\documentclass[preprint,aps]{revtex4-1}
\pdfoutput=1

\usepackage{amsmath,amssymb,amsfonts,dcolumn,color,graphicx,graphics,latexsym,placeins,epsfig}
\usepackage{epsfig}
\usepackage{bm}
\usepackage{slashed}
\usepackage{latexsym}
\usepackage{natbib}
\usepackage{url}
\usepackage{dcolumn}
\usepackage{color}
\usepackage{amsfonts,amssymb,amsmath}
\usepackage{graphicx,epsfig}
\usepackage{psfrag}
\usepackage{subfigure}
\usepackage{tabularx}
\usepackage{hyperref}
\hypersetup{colorlinks=true}
\usepackage{subcaption}

\newcommand{\be}{\begin{equation}}
\newcommand{\ee}{\end{equation}}
\newcommand{\ba}{\begin{eqnarray}}
\newcommand{\ea}{\end{eqnarray}}

\begin{document}

\title{Constraints on electrophilic scalar coupling from rotating magnetized stars and effects of cosmic neutrino background}
\author{Gaetano Lambiase$^{1,2}$\footnote{lambiase@sa.infn.it}}
\author{Tanmay Kumar Poddar$^{2}$\footnote{poddar@sa.infn.it}}
\affiliation{$^{1}$Dipartimento di Fisica E.R. Caianiello, Universit\`a di Salerno, Via Giovanni Paolo II 132 I-84084 Fisciano (SA), Italy, and \\
$^{2}$ INFN, Gruppo collegato di Salerno, Italy.\\}
\begin{abstract}
An ultralight electrophilic scalar field can produce a long-range Yukawa-like spatial profile around a rotating, magnetized star when coupled to the constant number density of electrons in either the magnetosphere or the star itself. This long-range scalar field generates an effective scalar charge in the star or its magnetosphere, leading to a long-range force between two compact stars in a binary system. The electrophilic scalar can also radiate from isolated pulsars or double pulsar binary systems. Using the Crab pulsar and PSR J0737-3039A/B as test cases, we derive constraints on the scalar-electron coupling by analyzing observations of long-range force, orbital period decay in binary systems, and pulsar spin-down rates. Among these, the most stringent limits on the coupling are obtained from the orbital period decay. However, these constraints can be significantly reduced if the scalar interacts with the pervasive cosmic neutrino background. Enhancing experimental sensitivity and studying compact objects with stronger magnetic fields and higher angular velocities could further strengthen these bounds.

\end{abstract}

\maketitle
\section{Introduction}
Rotating magnetized compact objects like Neutron Stars (NSs) and pulsars offer valuable opportunities to investigate new physics, including interactions between new degrees of freedom and Standard Model (SM) particles. These new degrees of freedom, which belong to Beyond Standard Model (BSM) physics, may or may not contribute to Dark Matter (DM) energy density in the universe. While direct detection experiments have imposed strict limits on Weakly Interacting Massive Particle (WIMP) DM candidates \cite{LUX:2013afz,PandaX-II:2017hlx,XENON:2018voc,XENON:2020gfr,Billard:2021uyg}, alternative candidates are being sought \cite{Hu:2000ke,Hall:2009bx,Hochberg:2014dra,Carr:2016drx}. One promising alternative is ultralight wave-like DM \cite{Hu:2000ke,Hui:2016ltb}, where DM behaves like a wave with a wavelength comparable to the size of a dwarf galaxy. Initially proposed to address small-scale structure challenges in the universe \cite{Marsh:2015wka,Robles:2018fur,Amin:2022pzv}, this form of DM is predominantly composed of scalar bosons with high occupation numbers, potentially exhibiting long-term oscillatory behavior or long-range effects.

Ultralight scalar DM can emerge from mechanisms like misalignment or the Stueckelberg mechanism \cite{Nelson:2011sf,Hui:2016ltb}, with its tiny mass explained by models such as clockwork or D-term inflation \cite{Fayet:1974jb,Fayet:2017pdp,Joshipura:2020ibd}. This type of DM interacts weakly with SM particles, and its coupling strength is constrained by numerous experiments and observations. Rigorous limits on scalar and vector couplings have been derived from neutrino oscillations \cite{Joshipura:2003jh,Bandyopadhyay:2006uh,Alonso-Alvarez:2021pgy,Alonso-Alvarez:2023tii,Gherghetta:2023myo}, neutrino decay \cite{Dror:2020fbh}, equivalence principle tests \cite{Wagner:2012ui}, magnetometer searches \cite{Kim:2021eye,Budker:2013hfa}, Gravitational Wave (GW) observations \cite{Arvanitaki:2014wva,Arvanitaki:2016qwi,Kopp:2018jom,KumarPoddar:2019ceq,KumarPoddar:2019jxe,Vermeulen:2021epa,Diamond:2023cto,Poddar:2023pfj,Lambiase:2023hpq}, atomic and nuclear transitions \cite{Campbell:2012zzb,Stadnik:2014tta,Hees:2018fpg}, pulsar timing arrays \cite{Kaplan:2022lmz}, Cosmic Microwave Background (CMB) measurements \cite{Hlozek:2014lca,Hlozek:2017zzf}, and many more \cite{AxionLimits,Nori:2018pka,KumarPoddar:2020kdz,Rogers:2020ltq,Poddar:2020qft,Tsai:2021irw,Tsai:2021lly,Poddar:2021ose,Poddar:2021sbc,Dentler:2021zij,Dalal:2022rmp,Bai:2023bbg,Poddar:2023bgk}.

Rotating NSs or pulsars act as cosmic beacons where electrons inside the stars or within the pulsar magnetosphere interact with a scalar field background. The scalar field need not be the DM candidate. These electrophilic scalar interactions can impact various observations, including orbital period decay in binary systems, long-range forces, and pulsar spin-down rates.


In scenarios where an ultralight electrophilic scalar field background interacts with a constant density distribution of Goldreich-Julian (GJ) electrons \cite{Goldreich:1969sb} in pulsar co-rotating magnetosphere or with electrons within the star, it generates a long-range scalar field profile around the pulsar magnetosphere or pulsar respectively. 

The electrophilic interaction of the long-range scalar field profile leads to a scalar-induced GJ charge in the magnetosphere and a scalar-induced net charge within the pulsar. These scalar charges gives rise to a long-range fifth force between two stars in a binary system, in addition to the gravitational force. 

The observed orbital period decay in binary systems, initially discovered in PSR B1913+16 (Hulse-Taylor binary) \cite{Hulse1975,Taylor1982,Weisberg:1984zz}, is mainly due to GW radiation. However, if the scalar field couples with GJ electrons in the pulsar magnetosphere or electrons within the pulsar, the radiation of this electrophilic scalar from the pulsar binary system could also contribute to the orbital period loss, alongside GW radiation.

The rotational energy loss of pulsars, leading to their spin-down, can be attributed to several factors. As highly magnetized NSs, pulsars experience a misalignment between their magnetic axis and rotation axis. This results in electromagnetic radiation, particularly magnetic dipole radiation emitted along the magnetic axis, which contributes to the pulsar's deceleration. Additionally, GW radiation from rapidly rotating pulsars also plays a role in the gradual decrease of the spin period. The interaction between the pulsar’s magnetic field and its surrounding nebula further accelerates this spin-down process. Moreover, the radiation of an ultralight scalar field interacting with the pulsar’s GJ charge could lead to scalar dipole radiation, which might also influence pulsar spin-down. Observations of this phenomenon provide an avenue to constrain such new interactions.

The scalar-electron interaction in both isolated pulsars and pulsar binaries can be suppressed in the presence of the ubiquitous Cosmic Neutrino Background (C$\nu$B). Similar to the CMB, the C$\nu$B permeates the universe. These relic neutrinos possess very low energies (on the order of $10^{-4}~\mathrm{eV}-10^{-6}~\mathrm{eV}$), making them challenging to detect. Due to their weak interactions, these low-energy C$\nu$B neutrinos could potentially provide information about the universe predating the CMB, if detected. Currently, the temperature of the C$\nu$B, with the CMB at $2.725~\mathrm{K}$, is about $1.95~\mathrm{K}$, corresponding to an energy of $1.68\times 10^{-4}~\mathrm{eV}$. The standard cosmological model predicts the number density of each flavor of cosmic Dirac neutrinos to be $56/\rm{cm^3}$, totaling $336/\rm{cm^3}$ when considering all flavors and their antiparticles. Depending on their masses, C$\nu$B neutrinos may be either relativistic or non-relativistic. Based on neutrino oscillation data \cite{ParticleDataGroup:2022pth}, two of the neutrino mass eigenstates are inferred to be non-relativistic today. Although detecting relic neutrinos is extremely difficult, several methods have been proposed. The PTOLEMY experiment aims to detect the C$\nu$B through the inverse beta decay of tritium \cite{Betts:2013uya}. Additionally, relic neutrinos could affect CMB fluctuations, providing an indirect means of detection \cite{Follin:2015hya}. Several other novel detection methods and indirect constraints have also been suggested for probing this elusive background \cite{Domcke:2017aqj,McKeen:2018xyz,Brdar:2022kpu,Arvanitaki:2022oby,Smirnov:2022sfo,Asteriadis:2022zmo}. The scalar mediated long-range force and the radiation from pulsar binaries or isolated pulsars can interact with the C$\nu$B medium, acquiring a medium-dependent scalar mass, similar to how photons acquire a plasma mass. This increase in scalar mass from its vacuum value weakens the constraints on the scalar coupling.

The paper is organized as follows. In Section \ref{newsec}, we explore the lepton content within rotating magnetized NS and their magnetospheres. In Section \ref{sec1}, we derive the long-range scalar field profile for the pulsar, resulting from the GJ charge in the pulsar magnetosphere and the net charge within the pulsar, assuming a constant density distribution. Section \ref{sec2} investigates the scalar mediated long-range force between binary stars with respect to gravity. Section \ref{sec4} provides calculations of massive scalar radiation from binary systems using a field-theoretical approach. In Section \ref{sec5}, we compute scalar dipole radiation from isolated pulsars, taking into account the influence of the GJ charge. The effect of the C$\nu$B on scalar-electron coupling is examined in Section \ref{sec6}. Section \ref{sec7} discusses the constraints on electrophilic scalar couplings derived from various observations. Finally, Section \ref{sec8} summarizes our results and presents the conclusions of the paper.

We use natural systems of units $(c=\hbar=1)$ throughout the paper unless stated otherwise.
\section{Leptons within the Neutron Star and its Magnetosphere}\label{newsec}
The distribution of neutrons, protons, electrons, and other particles such as muons and hyperons within a NS is influenced by the chosen Equation of State (EoS). The EoS relates pressure to density at different locations within the NS, resulting in non-uniform temperature distribution varying across distances within the star. For a typical NS density of $10^{15}~\rm{g/cm^3}$, the equilibrium compositions are neutrons, protons, electrons, muons and hyperons. There will be no muons and hyperons for density $10^{14}~\rm{g/cm^3}$ at low temperatures. For densities between $10^{6}~\rm{g/cm^3}$ and $10^{13}~\rm{g/cm^3}$, the stellar object contains electrons, neutrons and clusters (nuclei). Including the effect of nuclear clustering and nuclear forces, the NS models for a cold degenerate equation of states are found to be bound and stable within the density range $10^{14.08}$ to $10^{15.4}~\rm{g/cm^3}$ \cite{co01100t}. The threshold of muons is about $2.2\times 10^{14}~\rm{g/cm^3}$. Above the density $3\times 10^{14}~\rm{g/cm^3}$, the equation of state will be affected due to hyperons. Below the muon threshold, the number density of the muons is zero and we can write $\mu_n=\mu_e+\mu_p$, where $\mu$ denotes the chemical potential. We can also write $n_e=n_p$ from the charge conservation. Above the muon threshold, the equation for the chemical potential reads $\mu_n-\mu_p=\mu_e=\mu_\mu$ and the charge equality yields $n_p=n_e+n_\mu$. The number densities of electrons and other particles depend on the choice of mass density, pressure, and the adiabatic index. The number densities increase with increasing the mass density. For high mass densities, the electron and neutron numbers are roughly equivalent in magnitude. However, at lower mass densities, the electron number density is typically one to two orders of magnitude smaller than the neutron number density. Below a mass density of $2.2\times 10^{14}~\rm{g/cm^3}$, the muon number density is zero, but it increases as the mass density surpasses this threshold. In the low mass density regime (greater than muon threshold), the muon number density is approximately four orders of magnitude smaller than the neutron number density, while at higher mass densities, their magnitudes are more comparable. At high mass densities within NSs, the number densities of baryons and leptons are roughly equal in magnitude. However, at lower mass densities, the number densities differ by several orders of magnitude. Due to the varying density within the NS, electrons behave relativistically in the inner regions, while in the outer regions, they behave as non-relativistic particles. At the outer part of the NS, the mass density is lower, meaning the electrons are less densely packed, leading to a lower Fermi energy. Additionally, the temperature in the outer regions is relatively low, resulting in lower thermal energy for the electrons. Consequently, at the outer radius of the NS, the electrons exhibit non-relativistic behavior. The number of electrons in a NS can roughly vary from $10^{54}$ to $10^{57}$ depending on the NS density and to get stringent bound on the coupling, we choose the number of electron within the NS as $10^{57}$ in our following analysis. 

The number of electrons within the rotating NS is distinct from the number of electrons in the NS's magnetosphere that are co-rotating with the NS. These electrons are referred to as Goldreich-Julian (GJ) electrons, and their quantity depends on the pulsar's magnetic field and angular velocity. The general expression of effective number density for an electron is given as \cite{Babu:2019iml,Chauhan:2024qew}
\begin{equation}
\langle \bar{e}e\rangle = \tilde{\rho}=\int \frac{d^3p}{(2\pi)^3}\frac{m_e}{E_p}f(p),
\label{np1}
\end{equation}
where $f(p)$ denotes the phase space distribution function and Eq. \ref{np1} is different from the standard number density expression as 
\begin{equation}
\rho=\int \frac{d^3p}{(2\pi)^3}f(p).
\label{np2}
\end{equation}
For a non-relativistic particle, $E\sim m_e$ and Eq. \ref{np1} reduces to Eq. \ref{np2}. However, for relativistic distribution of particles, one should use Eq. \ref{np1}. The electrons satisfy Fermi-Dirac distribution function and if GJ electrons are non-relativistic, then $\tilde{\rho}=\rho=\rho^0_{\mathrm{GJ}}=2\Omega B/e$ (more discussion in Section \ref{sec1}). On the contrary, if GJ electrons are relativistic, then $\tilde{\rho}=\eta m_e(\rho^0_{\mathrm{GJ}})^\frac{2}{3}$,
where $\eta=\mathcal{O}(1)$, depends on the chemical potential and temperature of electrons. Considering Crab pulsar, which has a magnetic field of approximately $B\sim 8\times 10^{12}~\mathrm{G}$ and a spin period of $T=2\pi/\Omega\sim 33~\mathrm{ms}$, the number of relativistic co-rotating electrons is $N_\mathrm{GJ}\sim (m_e/2)\times(3\rho^0_{GJ}/\pi)^{2/3}\times (4/3)\pi R^3_{\mathrm{LC}}\sim 10^{44}$ and $(4/3)\pi R^3_{\mathrm{LC}}$ denotes the magnetosphere volume having radius equal to the Light Cylinder (LC) radius, $R_{\mathrm{LC}}\sim 1/\Omega$. We consider that the GJ electrons are relativistic within the LC. This is reasonable because the rotation of the NS generates strong electric fields in the magnetosphere, which accelerate the electrons to high velocities. The number of GJ electrons can be further enhanced by a factor as large as $\kappa\sim 10^6$ due to particle acceleration, pair production and plasma instability \cite{Lyutikov:2007fn,Kalapotharakos:2011vg,Timokhin:2015dua,Cruz:2020vfm}. However, $\kappa$ is not very well constrained and it can take different values for different pulsar magnetosphere models.
\section{Electrophilic Scalar field profile for a rotating magnetized neutron star}\label{sec1}
The current density for a rotating magnetized NS can be written as 
\begin{equation}
\mathbf{J}=\sigma(\mathbf{E}+\mathbf{v}\times \mathbf{B}),
\label{eq:ap9}
\end{equation} 
where $\sigma$ denotes the conductivity of the star and $\mathbf{v}$ denotes its velocity. Since the star is believed to be an excellent conductor, we can write $\frac{\mathbf{J}}{\sigma}\rightarrow 0$, and Eq. \ref{eq:ap9} reduces to
\begin{equation}
\mathbf{E}+\mathbf{\Omega}\times \mathbf{r}\times \mathbf{B}=0,
\label{eq:ap10}
\end{equation}
where $\mathbf{v}=\mathbf{\Omega}\times \mathbf{r}$, and $\mathbf{\Omega}$ denotes the angular velocity of the star. 
The free charges within the rotating conducting star will try to create a force-free equilibrium so that the net force on each charge becomes zero. To compensate for the Lorentz force, an electric field is created within the star. The strong electric field will create a volume charge density for a steady state configuration and is called the Goldreich-Julian (GJ) charge density \cite{Goldreich:1969sb}. Therefore, using Eq. \ref{eq:ap10} we obtain
\begin{equation}
\mathcal{J}_{GJ}(r)=\nabla\cdot \mathbf{E}\approx-2\mathbf{\Omega}\cdot \mathbf{B},
\label{eq:ap15}
\end{equation}
where the approximation holds for the velocity of the co-rotating charges $|\mathbf{v}|\ll1$, i.e., near the stellar surface. The tangential velocity of the charge increases as it moves away from the star. For a constant angular velocity, the speed of the charge cannot exceed the light speed to satisfy the causality condition. Therefore, from $\mathbf{v}=\mathbf{\Omega}\times \mathbf{r}$, we can write at $r=R_{\mathrm{LC}}$, $v=c$, and $R_{\mathrm{LC}}=1/\Omega$, where $R_{\mathrm{LC}}$ is called the Light Cylinder (LC) radius. Hence, charge particles with $r\leq R_{\mathrm{LC}}$ co-rotate with the star and form the magnetosphere of the pulsar. These charge particles are bound with the pulsar and cannot escape.

When charged particles, such as electrons, approach from the stellar surface to the light cylinder surface, their tangential velocity must increase to match the speed of light at that boundary. This increase in velocity occurs because the magnetic field lines extend outwards from the pulsar, and as the distance from the pulsar increases, the circumference of the circular path followed by the charged particles also increases. To maintain a constant rotational period, the velocity of the particles must increase as they move outward.

In the simplistic scenario, where $\mathbf{\Omega}$ and $\mathbf{B}$ are aligned, the GJ charge density can be expressed as $\rho^0_{GJ}=2\Omega B/e$. Additionally, we assume that the magnetic field remains constant throughout the LC. 
Therefore, we can write the scalar field Lagrangian, defining the interaction of a massive scalar field $\phi$ with the GJ electron number density as
\begin{equation}
\mathcal{L}=\frac{1}{2}\partial_\mu\phi\partial^\mu\phi-\frac{1}{2}m^2_\phi\phi^2-g_e\phi \rho_{GJ}.
\label{eq:a1}
\end{equation}
Thus, the equation of motion of the scalar field becomes
\begin{equation}
(\Box+m^2_\phi)\phi=g_e\rho_{GJ}.
\label{eq:a2}
\end{equation}
Considering the matter density distribution is spherically symmetric, we can write Eq. \ref{eq:a2} in radial coordinate as
\begin{equation}
\frac{\partial^2\phi}{\partial r^2}+\frac{2}{r}\frac{\partial\phi}{\partial r}-m^2_\phi\phi=-g_e\rho_{GJ}(r).
\label{eq:a3}
\end{equation}
The solution of the scalar field can be calculated from the variation of parameters method as
\begin{equation}
\phi(r)=\frac{g_e}{m_\phi r}\Big[e^{-m_\phi r}\int^r_0 r^\prime\rho_{GJ}(r^\prime)\sinh(m_\phi r^\prime)dr^\prime+\sinh(m_\phi r)\int^\infty_r r^\prime\rho_{GJ}(r^\prime)e^{-m_\phi r^\prime}dr^\prime\Big],
\label{eq:a4}
\end{equation}
where we follow \cite{Smirnov:2019cae,Babu:2019iml}. Assuming the GJ number density is constant and confined within $R_{\mathrm{LC}}$ of NS as 
\begin{eqnarray}
\rho_{GJ}(r) &=& \kappa \frac{m_e}{2}\Big(\frac{3}{\pi}\Big)^{\frac{2}{3}}\Big(\frac{2\Omega B}{e}\Big)^{\frac{2}{3}}, \hspace{0.2cm} r\leq R_{\mathrm{LC}}\nonumber\\
&=& 0,  \hspace{3.6 cm}r>R_{\mathrm{LC}}.
\end{eqnarray}

Imposing the boundary condition that the scalar field and its derivative are continuous at $r=R_{\mathrm{LC}}$, we obtain the solutions of the scalar field inside and outside of the star as
\begin{eqnarray}
\phi_\mathrm{GJ}(r)&=&\frac{g_e\kappa m_e}{2m^2_\phi}\Big(\frac{6\Omega B}{\pi e}\Big)^{\frac{2}{3}}\Big[-1+\frac{1+m_\phi R_{\mathrm{LC}}}{m_\phi r}e^{-m_\phi R_{\mathrm{LC}}}\sinh(m_\phi r)\Big],\hspace{1.5 cm} r\leq R_{\mathrm{LC}}\nonumber\\
&=& \frac{g_e\kappa m_e}{2 m^2_\phi}\Big(\frac{6\Omega B}{\pi e}\Big)^{\frac{2}{3}}\frac{e^{-m_\phi r}}{m_\phi r} \Big[\sinh(m_\phi R_{\mathrm{LC}})-m_\phi R_{\mathrm{LC}}\cosh(m_\phi R_{\mathrm{LC}})\Big],~~~r>R_{\mathrm{LC}}.
\label{eq:a5}
\end{eqnarray} 
In the small scalar mass limit, we can write Eq. \ref{eq:a5} as
\begin{eqnarray}
\phi_\mathrm{GJ}(r)&\approx & \frac{g_e \kappa m_e}{12}\Big(\frac{6\Omega B}{\pi e}\Big)^{\frac{2}{3}}(r^2-3R^2_{\mathrm{LC}}),~~~r\leq R_{\mathrm{LC}}\nonumber\\
&\approx & -\frac{g_e \kappa m_e R^3_{\mathrm{LC}}}{6r}\Big(\frac{6\Omega B}{\pi e}\Big)^{\frac{2}{3}},\hspace{1.5cm }r>R_{\mathrm{LC}}.
\label{eq:a6}
\end{eqnarray}
Thus, in the small scalar mass limit, the scalar field has a $1/r$ like long-range behaviour.

The scalar field can also couple with the net electron charge inside the NS. For simplicity, we assume that the electrons within the NS are non-relativistic. This assumption is well justified at the outer regions of the NS, where the mass density is lower, and both the Fermi energy and momentum of the electrons are small. In the inner regions of the NS, where the mass density is much higher, this assumption is less accurate. However, due to the lack of detailed knowledge about the star's internal properties, we conservatively assume that all electrons within the NS are non-relativistic. Additionally, stronger bounds on scalar-electron coupling would be obtained by considering non-relativistic electrons, as their number density is significantly larger than that of relativistic electrons. Therefore, we consider the constant number density profile for the net electrons within the NS as
\begin{eqnarray}
\rho_{N}(r) &=& \frac{3N}{4\pi R^3}, \hspace{0.2cm} r\leq R\nonumber\\
&=& 0,  \hspace{1 cm}r>R,
\end{eqnarray}
where the total number of electrons $N$ in the NS are taken to be non-relativistic. Similarly, imposing the boundary condition that the scalar field and its derivative are continuous at the surface of the star $(r=R)$, we obtain the scalar field solution as 
\begin{eqnarray}
\phi_N(r)&=&\frac{3g_e N}{4\pi R^3 m^2_\phi}\Big[-1+\frac{1+m_\phi R}{m_\phi r}e^{-m_\phi R}\sinh(m_\phi r)\Big],\hspace{1.5 cm} r\leq R\nonumber\\
&=& \frac{3g_e N}{4\pi R^3 m^2_\phi} \frac{e^{-m_\phi r}}{m_\phi r} \Big[\sinh(m_\phi R)-m_\phi R\cosh(m_\phi R)\Big],\hspace{0.9cm}r>R.
\label{eq:anj}
\end{eqnarray} 
In the limit of small scalar mass, Eq. \ref{eq:anj} reduces to
\begin{eqnarray}
\phi_N(r)&=&\frac{g_e N}{8\pi R^3}(r^2-3R^2),\hspace{1.5 cm} r\leq R\nonumber\\
&=& -\frac{g_eN}{4\pi r},\hspace{3.3 cm}r>R.
\label{eq:bnj}
\end{eqnarray} 

For a point source, the number density of the net electron charge can be written as $n(r)=N\delta^3(r)$ and the scalar field profile sourced by the NS is obtained as \cite{,Leefer:2016xfu,Dzuba:2024src}
\begin{equation}
\phi_N(r)=-\frac{g_eN}{4\pi r}e^{-m_\phi r}.
\label{eq:a7}
\end{equation}
\begin{figure}
\centering
\begin{minipage}{.5\textwidth}
  \centering
  \includegraphics[width=.9\linewidth]{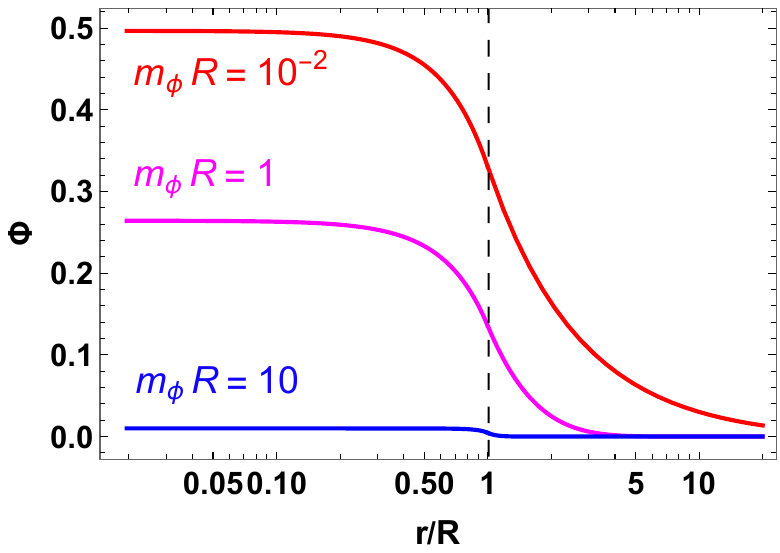}
  \captionof{figure}{$\Phi$ vs. $r$}
  \label{plot2a}
\end{minipage}%
\begin{minipage}{.5\textwidth}
  \centering
  \includegraphics[width=.9\linewidth]{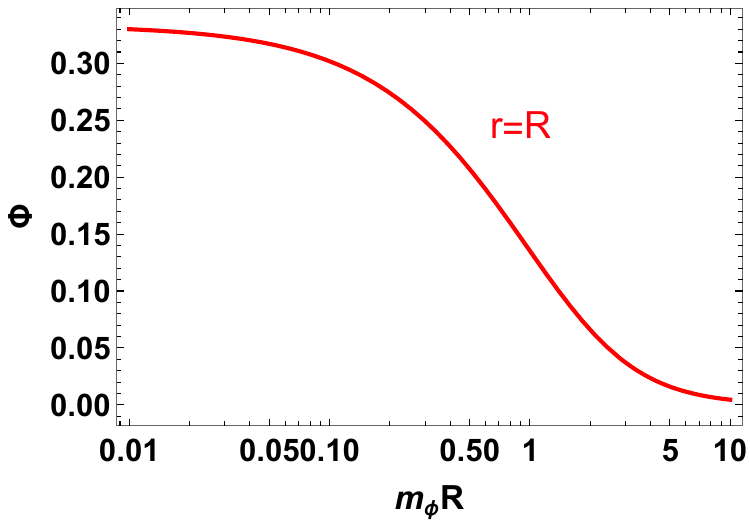}
  \captionof{figure}{$\Phi$ vs. $m_\phi$}
  \label{plot3a}
\end{minipage}
\end{figure}
In FIG. \ref{plot2a}, we obtain the variation of the scalar field profile $\Phi$ with $r$ sourced by the net electron charge within the star for $m_\phi R=10^{-2}$ (red), $1$ (magenta), $10$ (blue), using Eq. \ref{eq:anj}. Here, $\Phi(r)$ in the $y$ axis  is defined as $\Phi(r)=4\pi R\phi_N(r)/3g_eN$. For $m_\phi R\ll 1$, the scalar field profile shows a Coulomb potential-like behavior and for $m_\phi R\gg1$, the field $\Phi(r)$ essentially behaves as a step function. The $m_\phi\rightarrow 0$ limit (Eq. \ref{eq:bnj}), almost overlaps with the red curve. Outside of the star, $\phi(r)\rightarrow 0$ as $r\rightarrow \infty$.

In FIG. \ref{plot3a}, we obtain the variation of $\Phi(r)$ with respect to $m_\phi$ for $r=R$. The field value decreases with increasing $m_\phi R$, showing a long-range behavior. The mass of the scalar field sourced by the electrons within the NS is constrained by the radius of the star and for Crab pulsar, it results $m_\phi\lesssim 1/R\sim 1.41\times 10^{-11}~\mathrm{eV}$.

A similar behaviour of the scalar field profile as FIGs. \ref{plot2a} and \ref{plot3a} would arise from the scalar field sourced by the GJ charge. The mass of the long-range scalar field, sourced by the GJ charge, is constrained by the LC radius or the angular velocity of the NS. For Crab pulsar, $m_\phi\lesssim 1/R_{\mathrm{LC}}\sim \Omega\sim  1.22\times 10^{-13}~\mathrm{eV}$.

\section{Scalar mediated long-range force}\label{sec2}

We can write the scalar-induced GJ charge from Eq. \ref{eq:a5} as 
\begin{equation}
Q^{\mathrm{GJ}}_\phi=\frac{g_e\kappa m_e}{2 m^3_\phi}\Big(\frac{6\Omega B}{\pi e}\Big)^{\frac{2}{3}} \Big[\sinh(m_\phi R_{\mathrm{LC}})-m_\phi R_{\mathrm{LC}}\cosh(m_\phi R_{\mathrm{LC}})\Big].
\label{sw1}
\end{equation}
Similarly, the scalar induced net charge is obtained as
\begin{equation}
Q^N_\phi=\frac{3g_e N}{4\pi R^3 m^3_\phi} \Big[\sinh(m_\phi R)-m_\phi R\cosh(m_\phi R)\Big].
\label{sw2}
\end{equation}
These scalar induced charges sourced by the GJ electrons in the magnetosphere and net electrons within the NS result long-range force between two such stars in a binary system. The expression of the long-range force is given as 
\begin{equation}
F_\phi^{\mathrm{GJ}(N)}=\frac{Q^{\mathrm{GJ}(N)}_{\phi 1}Q^{\mathrm{GJ}(N)}_{\phi 2}}{r^2},
\label{sw3}
\end{equation}
where $r$ denotes the distance between two stars. This scalar-mediated long-range force, distinct from gravitational interaction, acts between two stars in a binary system. Such forces can be constrained through measurements from fifth-force searches and tests of the equivalence principle.

\section{Scalar field radiation from a binary system}\label{sec4}
In this section, we examine the energy loss caused by the radiation of a massive scalar field from a binary pulsar system. The scalar field can be sourced either by the electrons within the NS or by the GJ electrons in the magnetosphere. In \cite{Mohanty:1994yi}, the emission of massless scalar field radiation from a binary system is discussed using a field-theoretic approach. Meanwhile, the emission of corresponding massive scalar radiation from a multipole expansion (classical) method, considering a baryonic source density, is discussed in \cite{Krause:1994ar}. In this paper, we concentrate on calculating the emission of massive scalar particles from electronic current source, using a field-theoretic approach.

We begin by considering the scalar field being sourced by the electrons inside the star. We simplify the treatment by considering the stars as point sources, justified by the fact that the Compton wavelength of the scalar field $(1/\Omega_{\mathrm{orb}}\sim 10^9~\mathrm{km})$ is much larger than the dimensions of the stars $(R\sim 10~\mathrm{km})$. The Lagrangian describing the scalar field interaction with the electron number density can be expressed as
\begin{equation}
\mathcal{L}\supset g_e\phi n(r),
\label{ap1}
\end{equation}
where $n(r)=\sum_{j=1,2}N_j\delta^3(\mathbf{r}-\mathbf{r}_j(t))$ for a point source. $N_j$ stands for the total number of electrons within the j-th star and $\mathbf{r}_j$ denotes the position vector of the star in the centre of mass frame. Suppose, the motion of the binary stars is in the $x-y$ plane of a Keplerian orbit and its parametric form is given as \cite{Landau:1975pou}
\begin{equation}
x=a(\cos\xi-e), ~~~y=a\sqrt{1-e^2}\sin\xi, ~~~\Omega_{\mathrm{orb}} t=\xi-e\sin\xi,
\label{ap2}
\end{equation}
where $e$ and $a$ denote respectively the eccentricity and the semi-major axis of the Keplerian orbit and the fundamental orbital frequency is denoted as $\Omega_\mathrm{orb}=\sqrt{\frac{G(M_1+M_2)}{a^3}}$. Note, $\Omega_\mathrm{orb}$ is different from the spin angular frequency $\Omega$ of the star. Here, $M_1$ and $M_2$ designate the masses of the two stars. As the angular velocity is not constant for an elliptic orbit, we have to sum over all the harmonics of the fundamental frequency to calculate the scalar radiation. The coordinates in the frequency space are \cite{Landau:1975pou}
\begin{equation}
x(\omega)=\frac{aJ_n^\prime(ne)}{n}, ~~~y(\omega)=\frac{ia\sqrt{1-e^2}J_n(ne)}{ne},
\label{ap3}
\end{equation} 
where $\omega=n\Omega_{\mathrm{orb}}$ is the n-th harmonic of the fundamental frequency. The prime denotes the derivative of the Bessel function with respect to its argument. Therefore, we can write the emission rate of the scalar particles from the binary system as
\begin{equation}
d\Gamma=g^2_e|n(\omega^\prime)|^2 2\pi\delta(\omega-\omega^\prime)\frac{d^3k^\prime}{(2\pi)^3 2\omega^\prime}.
\label{ap4}
\end{equation}
The source number density in the frequency space is 
\begin{equation}
n(\omega)=\frac{1}{2\pi}\int\int e^{i\mathbf{k}\cdot\mathbf{r}}e^{-i\omega t}\sum_{j=1,2}N_j\delta^3(\mathbf{r}-\mathbf{r}_j(t))d^3r dt,
\label{ap6}
\end{equation}
which simplifies to 
\begin{equation}
n(\omega)=(N_1+N_2)\delta(\omega)+\Big(\frac{N_1}{M_1}-\frac{N_2}{M_2}\Big)M(ik_x x(\omega)+ik_yy(\omega))+\mathcal{O}(\mathbf{k}\cdot\mathbf{r})^2.
\label{ap7}
\end{equation}
Here $\mathbf{r_1}=\frac{M}{m_1}\mathbf{r}$ and $\mathbf{r_2}=-\frac{M}{m_2}\mathbf{r}$ in the centre of mass frame and $M$ is the reduced mass of the binary system consists of two stars with masses $M_1$ and $M_2$. Therefore, the rate of energy loss due to massive scalar radiation is 
\begin{equation}
\Big(\frac{dE_\phi}{dt}\Big)_N=\frac{g^2_e}{2\pi}\int|n(\omega^\prime)|^2\delta(\omega-\omega^\prime){\omega^\prime}^2d\omega^\prime\Big(1-\frac{m^2_\phi}{{\omega^\prime}^2}\Big)^\frac{1}{2},
\label{ap5}
\end{equation}
where the dispersion relation for the scalar particle is $\omega^2=k^2+m^2_\phi$.

Using Eqs. \ref{ap3} and \ref{ap6} we obtain the leading non zero contribution of $|n(\omega)|^2$ as
\begin{equation}
|n(\omega)|^2=\frac{1}{3}\Big(\frac{N_1}{M_1}-\frac{N_2}{M_2}\Big)^2M^2a^2\Omega^2_{\mathrm{orb}}\Big[{J^\prime_n{(ne)}}^2+\frac{1-e^2}{e^2}J_n(ne)^2\Big]\Big(1-\frac{m^2_\phi}{n^2\Omega^2_{\mathrm{orb}}}\Big),
\label{ap8}
\end{equation}
where we use the fact $\langle k^2_x\rangle=\langle k^2_y\rangle=\frac{k^2}{3}$. Therefore, substituting Eq. \ref{ap8} in Eq. \ref{ap5}, we obtain the rate of energy loss due to massive scalar radiation as
\begin{equation}
\Big(\frac{dE_\phi}{dt}\Big)_N=\frac{g^2_e}{6\pi}\Big(\frac{N_1}{M_1}-\frac{N_2}{M_2}\Big)^2M^2a^2\Omega^4_{\mathrm{orb}}\sum_{n>m_\phi/\Omega} n^2\Big[{J^\prime_n{(ne)}}^2+\frac{1-e^2}{e^2}J_n(ne)^2\Big]\Big(1-\frac{m^2_\phi}{n^2\Omega^2_{\mathrm{orb}}}\Big)^\frac{3}{2}.
\label{ap9}
\end{equation}
For massless scalar $(m_\phi\rightarrow 0)$, we obtain 
\begin{equation}
\Big(\frac{dE_\phi}{dt}\Big)_N=\frac{g^2_e}{12\pi}\Big(\frac{N_1}{M_1}-\frac{N_2}{M_2}\Big)^2M^2a^2\Omega^4_{\mathrm{orb}} \frac{\Big(1+\frac{e^2}{2}\Big)}{(1-e^2)^\frac{5}{2}},
\label{ap10}
\end{equation}
where we use the fact
\begin{equation}
\sum_{n} n^2\Big[{J^\prime_n{(ne)}}^2+\frac{1-e^2}{e^2}J_n(ne)^2\Big]=\sum_{n}f(n,e)=\frac{\Big(1+\frac{e^2}{2}\Big)}{2(1-e^2)^\frac{5}{2}}.
\end{equation}
The expression Eq. \ref{ap9} represents the massive scalar particle emission rate from a binary system where the scalar is coupled with the net electron charge within the star of the binary system. There should be an asymmetry between the charge-to-mass ratio of the two stars for the radiation to happen. The emission rate is dipolar and is proportional to $\Omega^4_{\mathrm{orb}}$. The rate of energy loss is only valid as long as $m_\phi<\Omega_{\mathrm{orb}}$ for the fundamental mode. The scalar emission rate is two times smaller than the equivalent vector emission rate \cite{KumarPoddar:2019ceq}.
\begin{figure}
\includegraphics[width=10cm]{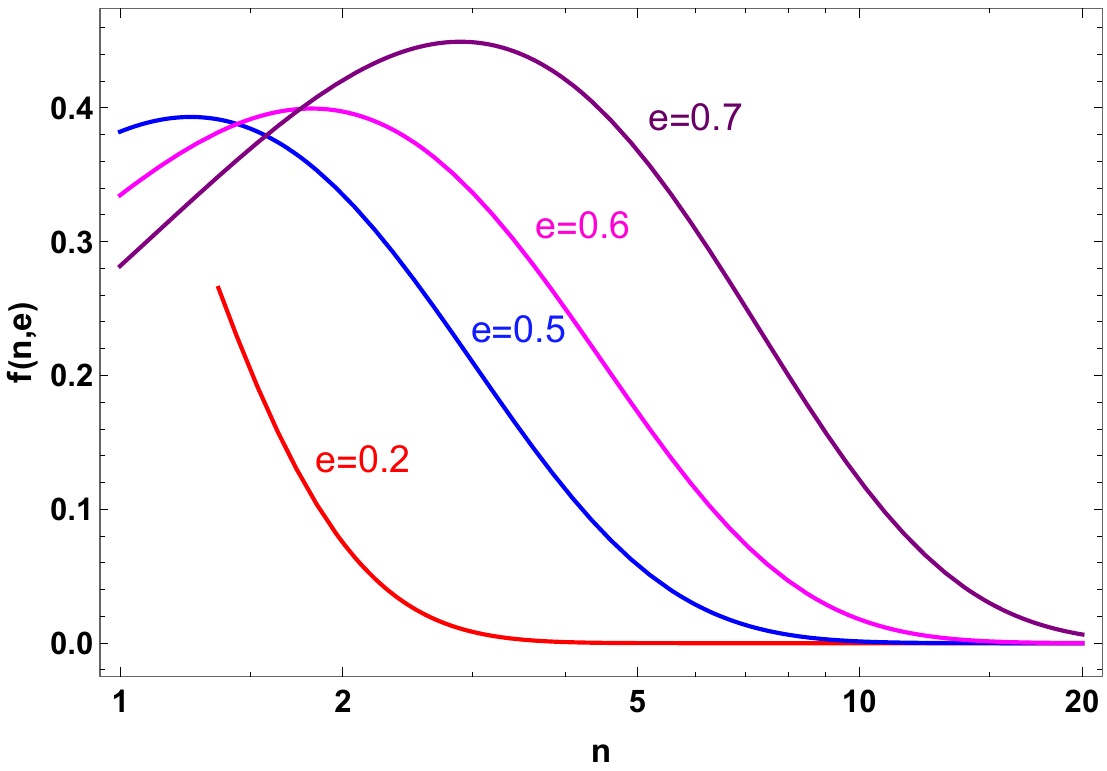}
\caption{Variation of $f(n,e)$ with $n$ for different eccentricity}
\label{plot1a}
\end{figure}

In FIG. \ref{plot1a} we plot the variation of $f(n,e)$ as a function of the number of harmonics $n$ for different orbital eccentricity values. The rate of energy loss increases with $f(n,e)$ and hence, eccentricity. The scalar radiation is dominated at higher harmonics as $e$ approaches one. For a fixed eccentricity value, the radiation spectrum has a peak at a particular value of $n$.

We can similarly obtain the scalar field radiation sourced by the GJ electrons. The number density of GJ electrons for a point source can be written as $n(r)=\sum_{j=1,2}N^{\mathrm{GJ}}_j\delta^3(\mathbf{r}-\mathbf{r}_j(t))$, where 
\begin{equation}
N^{\mathrm{GJ}}=\frac{2\pi\kappa m_e}{3\Omega^3} \Big(\frac{6\Omega B}{\pi e}\Big)^\frac{2}{3},
\label{gj1}
\end{equation}
considering $R_{\mathrm{LC}}\sim 1/\Omega$. The point source approximation is valid as $1/\Omega_{\mathrm{orb}}\gg R_{\mathrm{LC}}(1/\Omega)$, where $R_{\mathrm{LC}}\sim 10^3~\mathrm{km}$ for a typical millisecond pulsar. Similarly, we calculate the rate of energy loss due to scalar radiation sourced by GJ electrons as
\begin{equation}
\begin{split}
\Big(\frac{dE_\phi}{dt}\Big)_\mathrm{GJ}=\frac{2\pi^2\kappa^2m^2_eg^2_e}{27\pi}\Big(\frac{6}{\pi e}\Big)^{\frac{4}{3}}\Big[\frac{(\Omega_1B_1)^\frac{2}{3}}{\Omega^3_1M_1}-\frac{(\Omega_2B_2)^\frac{2}{3}}{\Omega^3_2M_2}\Big]^2M^2a^2\Omega^4_{\mathrm{orb}}\sum_{n>m_\phi/\Omega} n^2\Big[{J^\prime_n{(ne)}}^2+\\
\frac{1-e^2}{e^2}J_n(ne)^2\Big]\Big(1-\frac{m^2_\phi}{n^2\Omega^2_{\mathrm{orb}}}\Big)^\frac{3}{2}.
\end{split}
\label{gj2}
\end{equation}
Thus, we can have scalar radiation due to GJ charge even if the mass and radius of the two stars in the binary are the same, provided either or both of their surface magnetic fields and the spin frequencies are different. 


The rate of energy loss is related with the orbital period loss of the binary system as 
\begin{equation}
\dot{P_b}=-6\pi G^{-3/2}(M_1 M_2)^{-1}(M_1+M_2)^{-1/2}a^{5/2}\Big(\frac{dE_{GW}}{dt}+\frac{dE_{\phi}}{dt}\Big),
\label{orb_period}
\end{equation}
where $G$ denotes the gravitational constant and the energy loss due to the GW radiation is
\begin{equation}
\frac{dE_{GW}}{dt}=\frac{32}{5}G\Omega^6_{\mathrm{orb}} M^2a^4(1-e^2)^{-7/2}\Big(1+\frac{73}{24}e^2+\frac{37}{96}e^4\Big).
\label{gw}
\end{equation}
The orbital period of a binary pulsar system decreases due to GW radiation and matches with the general relativistic result to less than one per cent. The scalar radiation from the binary system can contribute to the orbital period loss of the binary system within the measurement uncertainty of the orbital period decay.
\section{Scalar field radiation from an isolated pulsar}\label{sec5}
In Section \ref{sec1}, we derived the long-range Yukawa-like scalar field profile for a time-independent constant density distribution of GJ electrons in the magnetosphere and electrons within the NS. In this section, we calculate the scalar radiation for a time-varying density distribution. Hence, the analysis in this section is entirely different from the previous discussion. Here, we consider two magnetosphere models—the co-rotating magnetosphere model and the polar cap model. In both these models, the magnetic moment axis and the rotation axis are not aligned and a small misalignment is necessary to get the pulsation effect.

We will now briefly summarize the derivation of the energy loss rate due to scalar emission, as previously shown in \cite{Krause:1994ar}. Suppose, the pulsar periodically rotates with an angular spin frequency $\Omega$ and the solution of the scalar field sourced by the pulsar can be written as a Fourier sum \cite{Krause:1994ar,Khelashvili:2024sup}
\begin{equation}
\phi(\mathbf{r},t)=\sum^{\infty}_{n=-\infty}e^{i\omega t}\phi_n(\mathbf{r}),
\label{r1}
\end{equation}
where $\omega=n\Omega$. The source current can also be written as a Fourier sum 
\begin{equation}
\rho(\mathbf{r},t)=\sum^{\infty}_{n=-\infty} e^{i\omega t} \rho_n(\mathbf{r}),
\label{r2}
\end{equation}
where 
\begin{equation}
\rho_n(\mathbf{r})=\frac{1}{T}\int^T_0 e^{i\omega t}\rho(\mathbf{r},t)dt.
\label{r3}
\end{equation}
Here, $T$ denotes the rotational time period of the pulsar and $T=2\pi/\Omega$. Therefore, the wave equation of the scalar field for each Fourier component is 
\begin{equation}
(\nabla^2+k^2)\phi_n(\mathbf{r})=-\rho_n(\mathbf{r}),
\label{r4}
\end{equation}
where $k^2=\omega^2-m^2_\phi$. We use the Green's function method to solve Eq. \ref{r4}. The solution of the Green's function for a point source is given as 
\begin{equation}
G(\mathbf{r},\mathbf{r^\prime})=\frac{1}{4\pi}\frac{e^{ik|\mathbf{r}-\mathbf{r^\prime}|}}{|\mathbf{r}-\mathbf{r^\prime}|}.
\label{r5}
\end{equation}
Thus, the solution of Eq. \ref{r4} becomes
\begin{equation}
\phi_n(\mathbf{r})=\frac{1}{4\pi}\int \frac{e^{ik|\mathbf{r}-\mathbf{r^\prime}|}}{|\mathbf{r}-\mathbf{r^\prime}|}\rho_n(\mathbf{r^\prime})d^3 r^\prime.
\label{r6}
\end{equation}
In the following we calculate the scalar radiation for the fundamental mode as the higher modes are already suppressed. In the limit $|\mathbf{r}|\gg |\mathbf{r^\prime}|$ and $k|\mathbf{r^\prime}|\ll 1$, we can write the scalar field solution as
\begin{equation}
\phi_{\pm}(r)=\frac{e^{\pm ikr}}{4\pi r}(Q_{\pm}\pm ik\hat{\mathbf{n}}\cdot\mathbf{p}_{\pm}+...),
\label{r7}
\end{equation}
as the number of harmonics $(n)$ can take both positive and negative values. The monopole and dipole moments are 
\begin{equation}
Q_{\pm}=\int d^3r^\prime \rho(\mathbf{r}^\prime),~~~\mathbf{p}_{\pm}=\int d^3r^\prime\rho({\mathbf{r^\prime}})\mathbf{r^\prime,}
\label{r8}
\end{equation}
respectively. For a source of conserved charge, the monopole term vanishes. Therefore, the outgoing dipolar scalar field solution can be written as \cite{Krause:1994ar}
\begin{equation}
\phi(\mathbf{r},t)=\frac{ik}{4\pi r}(\hat{\mathbf{n}}\cdot \mathbf{p}_+e^{-i(\Omega t-kr)}-\hat{\mathbf{n}}\cdot \mathbf{p}_-e^{i(\Omega t-kr)}).
\label{r9}
\end{equation}
Thus, the rate of energy loss due to the scalar radiation is
\begin{equation}
\frac{dE}{dt}=\int r^2d\Omega_n (\hat{\mathbf{n}}\cdot \mathbf{S}),
\end{equation}
where the energy flux is $\mathbf{S}=\dot{\phi}^*\nabla\phi$ and $d\Omega_n$ is the solid angle in the $\theta-\phi$ plane. Taking the time average over the rotation period, we obtain
\begin{equation}
\frac{dE}{dt}=\frac{1}{8\pi^2}\Omega k^3\int d\Omega_n|\mathbf{p_+}\cdot \hat{\mathbf{n}}|^2,
\label{osc}
\end{equation}
as $\mathbf{p_+}=\mathbf{p^*_-}$ and the integration has to be done over the solid angle. The Fourier component of the dipole moment is
\begin{equation}
\mathbf{p}_+=\frac{1}{T}\int e^{i\Omega t}dt\int d^3r^\prime\rho(\mathbf{r}^\prime,t)\mathbf{r}^\prime.
\label{at}
\end{equation}
In the following, we calculate the rate of energy loss due to scalar dipole radiation in the two above-mentioned magnetosphere models.

\subsection{Corotating magnetosphere model}
The GJ electrons can source a scalar-induced dipole moment and contribute to the pulsar spin-down. The magnetic colatitude $\theta_m$ is defined as $\cos\theta_m=\cos\alpha\cos\theta+\sin\alpha\sin\theta\cos(\phi-\Omega t)$ \cite{Melrose:2011eh}, where $\alpha$ denotes the angle between the magnetic moment axis and the spin axis. The initial condition is chosen as $\phi=0$ at $t=0$, corresponds to the fact that the magnetic axis is in the plane. The value of $\theta_m$ varies periodically as the star rotates. The field line is in an azimuthal plane at the phase values $\phi-\omega t=n\pi$, where $n=0,\pm 1,\pm 2,...$. For simplicity, we assume that the angular velocity of the NS has only an azimuthal component, and thus, the azimuthal component of the magnetic field contributes to the number density. The azimuthal component of the magnetic field is given as \cite{Melrose:2011eh}
\begin{equation}
B_\theta(r)=\frac{B_0R^3}{r^3}[\cos\alpha\sin\theta-\sin\alpha\cos\theta\cos(\phi-\Omega t)],
\label{mag1}
\end{equation}
where $B_0$ denotes the surface magnetic field and $R$ denotes the radius of the star as mentioned earlier. We can also write the equation of motion for the scalar field as 
\begin{equation}
(\Box+m_\phi^2)\phi=g_e \Big(\frac{2\mathbf{\Omega}\cdot \mathbf{B}}{e}\Big),
\label{mag2}
\end{equation}
where the scalar-induced number density for a dipolar magnetic field can be obtained as
\begin{equation}
\rho (r,t)\approx \frac{2g_eB_0R^3\Omega}{er^3}\tan\alpha\cos\theta_m\cos(\phi-\Omega t),
\label{t1}
\end{equation}
where we omit the time-independent terms as they will not contribute to the radiation and remove other subleading terms because of small $\alpha$. The region of interest in the radial direction is from $R$ to $x R$, where $x$ is slightly greater than one and we are confining ourselves near the surface of the star. Also, near the surface of the star, the electrons can be taken as non-relativistic. The speed of the Crab pulsar, and consequently, the velocity of charged particles on its surface, are computed as $8.6\times 10^{-3}\ll 1$, which can be approximately treated as non-relativistic. Note, we are not extending the radial limit to $R_{\mathrm{LC}}$, as near the light cylinder, the magnetic field behaviour largely deviates from the dipolar nature. Hence, the radiation happens at angular frequency $\Omega$ with scalar-induced dipole moment 
\begin{equation}
|\mathbf{p}|=\int d^3r^\prime \mathbf{r^\prime}\rho(r^\prime)\approx \frac{\pi^2}{e}g_e B_0 x R^4\Omega \tan\alpha\cos\theta_m.
\label{t2}
\end{equation}
Using Eqs. \ref{osc} and \ref{t2} we obtain the rate of energy loss for the pulsar spin-down due to scalar radiation as
\begin{equation}
\frac{dE}{dt}=\frac{1}{8\pi^2}\Omega k^3\int\int d\Omega_n |\hat{\mathbf{n}}\cdot \mathbf{p}_\Omega|^2\approx\frac{\pi^3}{8e^2}g^2_eB^2_0x^2 R^8\Omega^6\sin^2\theta_m\Big(1-\frac{m^2_\phi}{\Omega^2}\Big)^\frac{3}{2},
\label{t3}
\end{equation}
where we use the fact $\alpha\sim\theta_m$ for small $\alpha$. Also, in this case, the radiation only happens for $\Omega>m_\phi$. 

The expression Eq. \ref{t3} is valid for $n=1$ Fourier mode and corresponds to the case where the pulsar is rotating with $\Omega$ fundamental spin frequency. The $n>1$ modes are highly suppressed by powers of $v\sim \Omega R\ll1$. The pulsar spin-down due to the scalar radiation is valid only for $\Omega>m_\phi$. The radiation of the scalar field can contribute to the pulsar spin-down within the uncertainties arises due to different EoS, timing model, parallax method etc.
\subsection{Polar cap model}
We now focus on the Polar Cap (PC) model \cite{Ruderman:1975ju,Timokhin:2015dua,Timokhin:2018vdn,Cruz:2020vfm,Caputo:2023cpv}, where the magnetosphere is mostly screened, and the scalar field is sourced only in the PC regions, just above the magnetic poles. In a screened magnetosphere, electrons flow along the open magnetic field lines from the polar cap region. We assume that the scalar's mass is smaller than the NS's spin frequency, allowing the PC region to be treated as a point particle with a scalar-induced charge
\begin{equation}
Q_\phi=g_e\int_{\mathrm{PC}} d^3x \Big\langle\frac{2\mathbf{\Omega}\cdot \mathbf{B}}{e}\Big\rangle_t,
\label{pc1}
\end{equation}
where $\langle\cdot\rangle$ represents the time average and we model the polar cap as a cylinder with radius $r_{\mathrm{PC}}=R\sqrt{\Omega R}$ and height $h\sim 7~\mathrm{m}(\Omega/(2\pi\times 30~\mathrm{Hz}))^{-4/7}(B/(8.5\times 10^{12}~\mathrm{G}))^{-4/7}$, which is a conservative value. The actual value of $h$ depends on the number of charged particles in the PC region \cite{Khelashvili:2024sup}. Therefore, Eq. \ref{pc1} becomes $Q_\phi\sim g_e\pi r^2_{\mathrm{PC}}h(2\Omega B_0/e)$, where we take the surface values for $\Omega$ and $B$. Also, in the polar cap model, $r_{\mathrm{PC}}\ll R\ll 1/\Omega$ and hence, the two polar caps at the north and south magnetic poles can be treated as point sources rotating like a dipole. We model the two polar caps have equal and opposite charges $Q_\phi^N=-Q_\phi^S$, velocities $v^N_{\mathrm{PC}}=-v^S_{\mathrm{PC}}$ and positions $r_{\mathrm{PC}}^N=-r_{\mathrm{PC}}^S=R(t)$. This is justified since the dipolar magnetic field points outward from the star at the north pole and inward at the south pole. Therefore, $\mathbf{\Omega}\cdot \mathbf{B}$ has different signs in the two poles. Here, $N$ and $S$ denote northern and southern poles respectively. Therefore, we can write the scalar-induced charge density for a rotating dipole in the PC model as
\begin{equation}
\rho(r,t)=Q_\phi(\delta^3(\mathbf{r}-\mathbf{R}(t))-\delta^3(\mathbf{r}+\mathbf{R}(t))),
\label{scalar1}
\end{equation}
where
\begin{equation}
\mathbf{R}(t)=R(\sin\theta_m\cos(\Omega t)\hat{\mathbf{x}}+\sin\theta_m\sin(\Omega t)\hat{\mathbf{y}}+\cos(\Omega t)\hat{\mathbf{z}}).
\label{scalar2}
\end{equation}
Using Eqs. \ref{at}, \ref{scalar1}, and \ref{scalar2} we write the Fourier component of the dipole moment as 
\begin{equation}
\hat{\mathbf{n}}\cdot \mathbf{p}_\Omega=Q_\phi R\sin\theta_m\sin\theta_n e^{i\phi_n},
\label{scalar3}
\end{equation}
where $\theta_n$ and $\phi_n$ denote the spherical coordinates with respect to the observer and $\hat{\mathbf{n}}$ denotes the unit vector along $\mathbf{r}$. Hence, the rate of energy loss due to dipole scalar radiation is 
\begin{equation}
\frac{dE}{dt}=\frac{1}{3\pi}Q^2_\phi R^2\Omega^4\sin^2\theta_m\Big(1-\frac{m^2_\phi}{\Omega^2}\Big)^\frac{3}{2}\approx \frac{4\pi}{3e^2}g^2_e\Omega^8R^8B^2_0h^2\sin^2\theta_m\Big(1-\frac{m^2_\phi}{\Omega^2}\Big)^\frac{3}{2},
\label{scalar4}
\end{equation}
where we use Eqs. \ref{osc}, \ref{scalar3} and the expression for $Q_\phi$. Also, in the PC model, the scalar radiation happens for $\Omega>m_\phi$.
\section{Effects of cosmic neutrino background on scalar field profile}\label{sec6} 
The mass of the ultralight scalar particle increases when it propagates through the ubiquitous C$\nu$B medium. Depending on the mass of neutrinos, the background can be either relativistic or non-relativistic. The thermal mass of the scalar transforms as \cite{Babu:2019iml,Chauhan:2024qew}
\begin{equation}
m^2_\phi\rightarrow m^2_\phi+y^2_\nu\frac{n_\nu}{m_\nu},
\label{cnb1}
\end{equation}
where $y_\nu$ denotes the scalar-C$\nu$B coupling, $m_\nu$ denotes the neutrino mass and $n_\nu$ denotes the C$\nu$B density, assuming non-relativistic neutrinos. Thus, the thermal correction of the scalar mass $\Delta m^2_\phi=y^2_\nu\frac{n_\nu}{m_\nu}$ can have significant contribution on the scalar field profile due to the smallness of neutrino mass. For relativistic neutrinos, the number density modifies as $n_\nu\rightarrow \alpha m_\nu n^{2/3}_\nu $, where $\alpha\sim \mathcal{O}(1)$, depends on the neutrino momentum distribution function. The relativistic neutrino density is lowered by a factor of $m_\nu/E_\nu$ compared to the non-relativistic spectrum. The mass correction $\Delta m^2_\phi$ for the non relativistic C$\nu$B $(m_\nu\gg 1.7\times 10^{-4}~\mathrm{eV} \sim 1.95~\mathrm{K})$ can be written as \cite{Chauhan:2024qew}
\begin{equation}
\Delta m^2_\phi=y^2_\nu\frac{n_\nu}{m_\nu}\sim 10^{-32}~\mathrm{eV}^2\Big(\frac{y_\nu}{10^{-10}}\Big)^2\Big(\frac{n_\nu}{56/\mathrm{cm}^3}\Big)\Big(\frac{0.1~\mathrm{eV}}{m_\nu}\Big),
\label{cnb3}
\end{equation} 
whereas for relativistic neutrinos \cite{Chauhan:2024qew}
\begin{equation}
\Delta m^2_\phi=\alpha y^2_\nu n_\nu^\frac{2}{3}\sim 10^{-29}~\mathrm{eV}^2\Big(\frac{y_\nu}{10^{-10}}\Big)^2\Big(\frac{n_\nu}{56/\mathrm{cm}^3}\Big)^{\frac{2}{3}}.
\label{cnb4}
\end{equation}
Based on oscillation data, it can be inferred that two of the three neutrino mass eigenstates are non-relativistic today, while the remaining mass eigenstate may still be relativistic. 
The cosmic neutrino density can affect the scalar field profile and hence, screen the bounds of coupling parameters. As the mass of the scalar increases in presence of the C$\nu$B medium, the long-range force becomes effectively short-ranged.
\section{Constraints from observations}\label{sec7}
In the following, we obtain constraints on the scalar-electron coupling from the orbital period loss from double pulsar binary, search for fifth force, and pulsar spin-down. We also investigate how these derived bounds quench due to the presence of C$\nu$B. The derived constraints depend on the source currents. The source can be either the GJ charge density in the magnetosphere or the net charge density within the NS. Several bounds on the scalar-electron coupling has already been obtained from fifth force, astrophysics and DM search experiments. The Eot-Wash experiment, which tests the equivalence principle, places a constraint on the scalar-electron coupling as $g_e\lesssim 10^{-24}$ for $m_\phi\lesssim 10^{-14}~\mathrm{eV}$ \cite{Hees:2018fpg}. The MICROSCOPE experiment, also designed to test the equivalence principle, sets a constraint on the scalar-electron coupling as $g_e\lesssim 10^{-25}$ for $m_\phi\lesssim 10^{-14}~\mathrm{eV}$ \cite{Berge:2017ovy}. Fifth force experiments, which test the inverse square law, also place constraints on the scalar-electron coupling \cite{Fischbach:1996eq,Berge:2017ovy,KONOPLIV2011401}. In the low-mass limit, the Yukawa-type fifth force simplifies to the Newtonian gravitational force law. The energy loss rate of the White Dwarf (WD) puts the bound on scalar-electron coupling as $g_e\lesssim 10^{-16}$ in the ultralight scalar mass limit \cite{Bottaro:2023gep}. Several DM search experiments can also place bounds on $g_e$. The cryogenic resonant mass AURIGA detector, which can detect oscillations of solid bodies caused by interactions with a scalar DM halo, places constraints on the scalar-electron coupling as $g_e\lesssim 10^{-27}$ for $m_\phi\sim 10^{-12}~\mathrm{eV}$ \cite{Branca:2016rez}. The space-time separated atomic clocks and cavities put bound on the electrophilic oscillating scalar DM coupling for the scalar mass $10^{-19}~\mathrm{eV}\lesssim m_\phi\lesssim 10^{-15}~\mathrm{eV}$ \cite{Filzinger:2023qqh}. The optical spectroscopy measurements (Cs/Cav) puts bound on the scalar-electron coupling for scalar mass $10^{-10}~\mathrm{eV}\lesssim m_\phi\lesssim 10^{-7}~\mathrm{eV}$ \cite{Tretiak:2022ndx}. The DArk Matter from Non Equal Delays (DAMNED) interferometer puts bound on the scalar DM-electron coupling for the scalar mass $10^{-11}~\mathrm{eV}\lesssim m_\phi\lesssim  10^{-9}~\mathrm{eV}$ \cite{Savalle:2020vgz}. The gravitational wave detector, GEO 600 interferometer is used to put bound on $g_e$ by measuring the optical length variation of the interferometer due to oscillating scalar DM interaction. The bound is obtained for the scalar mass $10^{-13}~\mathrm{eV}\lesssim m_\phi\lesssim 10^{-11}~\mathrm{eV}$ \cite{Vermeulen:2021epa}. Time oscillating sub-eV scalar DM induces apparent oscillation in the fundamental constant which results oscillation in the size and index of refraction of the solid. Using cross-correlated data of Fermilab Holometer, bound on scalar-electron coupling has been obtained for scalar mass $10^{-12}~\mathrm{eV}\lesssim m_\phi\lesssim 10^{-7}~\mathrm{eV}$ \cite{Aiello:2021wlp}. The variation of fundamental constant due to time oscillating scalar DM from cryogenic Sapphire oscillator, Hydrogen Maser atomic oscillator, and bulk acoustic wave quartz oscillator yield bound on the scalar-electron coupling for the scalar mass $10^{-19}~\mathrm{eV}\lesssim m_\phi\lesssim 10^{-14}~\mathrm{eV}$ \cite{Campbell:2020fvq}. The Iodine spectroscopy experiment puts bound on the scalar-electron coupling for the scalar mass $10^{-13}~\mathrm{eV}\lesssim m_\phi\lesssim 10^{-7}~\mathrm{eV}$ \cite{Oswald:2021vtc}. The frequency comparisons between strontium optical lattice clock, cryogenic crystalline silicon cavity, and a hydrogen maser due to time variation in electron mass induced by interaction with the scalar DM puts bound on the coupling for scalar mass $10^{-21}~\mathrm{eV}\lesssim m_\phi\lesssim 10^{-16}~\mathrm{eV}$ \cite{Kennedy:2020bac}. The frequency comparisons of a quartz oscillator to that of hyperfine and electronic transitions of Rubidium and Dysprosium due to scalar DM-electron interaction puts bound on the coupling for the scalar mass $10^{-17}~\mathrm{eV}\lesssim m_\phi\lesssim 10^{-13}~\mathrm{eV}$ \cite{Zhang:2022ewz}. Similarly, the frequency comparisons of $^{171}\mathrm{Yb}$ optical lattice clock and $^{133}\mathrm{Cs}$ fountain clock yield bound on the scalar DM-electron coupling for the scalar mass $10^{-22}~\mathrm{eV}\lesssim m_\phi\lesssim 10^{-20}~\mathrm{eV}$ \cite{Kobayashi:2022vsf}. The LIGO O3 result also puts bound on the scalar-electron coupling from the oscillation of interferometer's beamsplitter and arm test masses for the scalar mass $10^{-14}~\mathrm{eV}\lesssim m_\phi\lesssim 10^{-11}~\mathrm{eV}$ \cite{Gottel:2024cfj}. The NANOGrav data also puts bound on the scalar-electron coupling for the scalar mass $10^{-24}~\mathrm{eV}\lesssim m_\phi\lesssim 10^{-21}~\mathrm{eV}$ \cite{NANOGrav:2023hvm}. In the following, we obtain bounds on the scalar-electron coupling by observing rotating magnetized NS and compare with the existing constraints.

\subsection{Orbital period loss of a double pulsar binary}

\begin{figure}
\includegraphics[width=12cm]{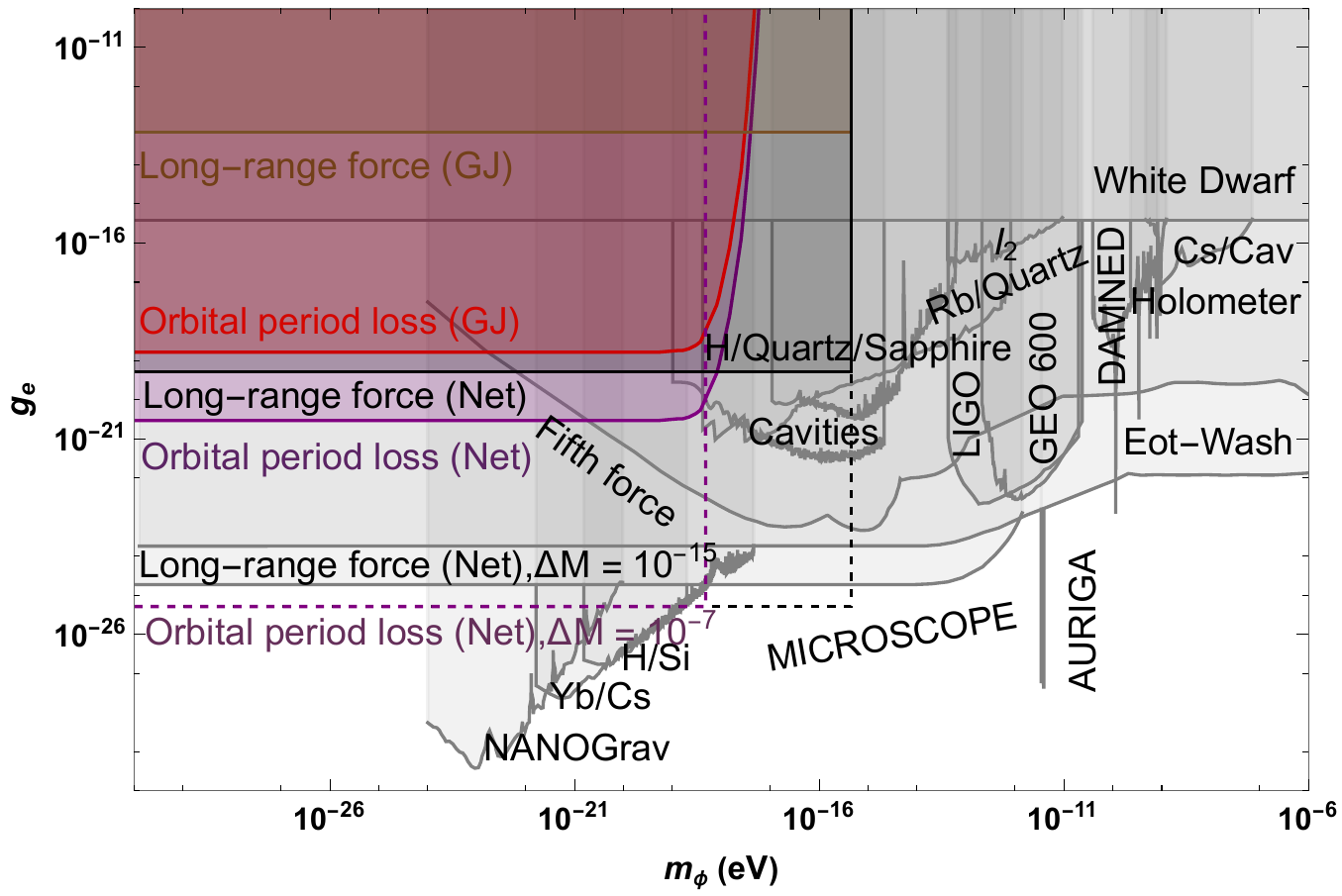}
\caption{Constraints on $g_e$ from the orbital period loss of PSR J0737-3039A/B, and search for long range force. The gray shaded regions are excluded from existing limits and the coloured shaded regions are our results.}
\label{plotnew}
\end{figure}

The orbital period of a binary system decreases due to GW radiation which matches with Einstein's general relativistic result to less than one per cent. The long range scalar field can couple with GJ electrons in the pulsar magnetosphere and electrons within the star. The scalar radiation from double pulsar binary due to such coupling can contribute in the orbital period loss of the system. The contribution of the scalar radiation should be within the measurement uncertainty of the orbital period decay. We consider PSR J0737-3039A/B \cite{Kramer:2006nb} as a test pulsar binary system to derive the scalar coupling. The masses of the two stars in the binary are $M_1=1.3381\pm 0.0007~M_\odot$ and $M_2=1.2489\pm 0.0007~M_\odot$, the eccentricity is $e=0.087$, the orbital frequency is $\Omega_\mathrm{orb}=4.79\times 10^{-19}~\mathrm{eV}$, and the semi major axis is $a=4.83\times 10^{15}~\mathrm{eV}^{-1}$. The surface magnetic fields of the two stars are $B_1\sim 6.3\times 10^9~\mathrm{G}$ and $B_2\sim 1.2\times 10^{12}~\mathrm{G}$. Their spin frequencies are $\Omega_1\sim 44.054~\mathrm{Hz}$ and $\Omega_2\sim 0.361~\mathrm{Hz}$. As the masses of the two stars in the binary are different, $(N_1/M_1-N_2/M_2)\neq 0$, even for $N_1=N_2$, where $N_1$ and $N_2$ denote the net electrons within the two pulsars. Therefore, the scalar radiation contribution can be non-zero. The difference in the number-to-mass ratio of the two stars can also be treated as follows. We can write $N_i(m_p+m_e)\approx N_i m_p=M_i-GM^2_i/R_i$, where $GM^2_i/R_i$ is the gravitational binding energy of the i-th star and $m_p$ is the mass of the proton. Therefore, one can write $(N_1/M_1-N_2/M_2)=G/m_p(M_2/R_2-M_1/R_1)$. We use $N_1=N_2\sim 10^{57}$ to derive the bounds. For scalar field radiation coupled with GJ electrons, radiation is possible for equal-mass pulsars, as long as their magnetic fields or angular frequencies are different.

In FIG. \ref{plotnew}, the purple and red shaded regions are excluded from the orbital period loss of PSR J0737-3039A/B due to scalar radiation coupled with net electrons within the star and GJ electrons in the magnetosphere respectively. We obtain the bounds on the scalar-electron coupling as $g_e\lesssim 3\times 10^{-21}$ for electrons within the star and $g_e\lesssim 7.24\times 10^{-19}$ for GJ electrons in the magnetosphere. We use Eqs. \ref{ap9}, \ref{gj2}, \ref{orb_period}, and \ref{gw} in obtaining the bounds. These bounds are valid for the mass of the scalar $m_\phi\lesssim 4.79\times 10^{-19}~\mathrm{eV}$, constrained by the orbital frequency of the binary system.

The bound on the scalar coupling obtained from the net electron charge within the star is five orders of magnitude more stringent than the bound from WD. It is also stronger than the fifth force constraints in the lower scalar mass limit, though it is four orders of magnitude less stringent than the MICROSCOPE bound. Stars with greater mass, larger semi-major axis, higher angular frequencies, magnetic fields and experiments with better sensitivities can lead to stronger constraints. For instance, if the masses of the binary stars are measured with an accuracy of $\Delta M=10^{-7}$ (purple dashed line), the resulting bound on the orbital period decay originating from the net charge would surpass the existing constraint.

Similar to scalar radiation, ultralight vector particles can also be emitted from a binary system, contributing to orbital period loss \cite{Krause:1994ar, KumarPoddar:2019ceq}. In the massless limit, the energy loss due to vector radiation is twice that of scalar radiation. Ultralight scalar particles can also couple to the muons inside the NS \cite{Garani:2019fpa}, and muonphilic coupling can be constrained in a similar manner by analyzing the orbital period loss in binary systems \cite{KumarPoddar:2019ceq}.

\subsection{Search for long range force}
\label{newsubsec}
The scalar-induced charge, sourced by GJ electrons in the magnetosphere and electrons within the star, can lead to a long-range force between two stars in a binary system. In the small scalar mass limit, the scalar charge induced by the GJ electrons is expressed as
\begin{equation}
 Q^{\mathrm{GJ}}_\phi=\frac{g_e\kappa m_e R^3_{\mathrm{LC}}}{6}\Big(\frac{6\Omega B}{\pi e}\Big)^\frac{2}{3}, 
 \label{news1}
\end{equation}
and in the small scalar mass limit, the scalar-induced charge sourced by the electrons within the star is given as
\begin{equation}
    Q^\mathrm{N}_\phi=\frac{g_eN}{4\pi}.
    \label{news2}
\end{equation}
Now, we consider the double pulsar binary system PSR J0737-3039A/B to measure the long-range fifth force between the two pulsars in the binary. The ratio of the scalar-mediated fifth force to the gravitational force between the two stars is given as
\begin{equation}
    \varepsilon=\frac{F_\phi^{\mathrm{GJ}/\mathrm{N}}}{F_G}=\frac{Q_{\phi 1}^{\mathrm{GJ}/\mathrm{N}}Q_{\phi 2}^{\mathrm{GJ}/\mathrm{N}}}{G M_1 M_2}.
    \label{news3}
\end{equation}
The uncertainty in the mass measurement of the stars in the binary system PSR J0737-3039A/B is approximately $7\times 10^{-4}$ \cite{Kramer:2006nb}. Assuming that the contribution from the scalar-mediated long-range force remains within this measurement uncertainty, we derive conservative bounds on the scalar-electron coupling: $g_e\lesssim 6.79\times 10^{-14}$ for GJ electrons in the magnetosphere and $g_e\lesssim 5.14\times 10^{-20}$ for electrons within the star. In FIG. \ref{plotnew}, the brown and black shaded regions denote the exclusion of the scalar-electron coupling for searching the scalar mediated force sourced by the GJ electrons and net electrons within the star respectively. The bounds are only valid for the mass of the scalar $m_\phi\lesssim 4.5\times 10^{-16}~\mathrm{eV}$, constrained by the distance between the two stars in the binary. The bound on the coupling due to electrons can be improved by examining binary systems with pulsars or magnetars that have stronger magnetic fields and measurements with better sensitivities. For example, if the mass measurement accuracy reaches $\Delta M=10^{-15}$,  the bound on the coupling derived from the long-range force generated by the net charge becomes significantly more stringent compared to existing constraints.
\subsection{Search from pulsar spin-down}
\label{subsecd}
Pulsars gradually lose rotational energy through electromagnetic radiation, magnetic dipole radiation, and GWs, leading to a reduction in their spin rate over time. The spin-down luminosity of the Crab pulsar has been measured as $L=4.5\times 10^{38}~\mathrm{erg}/\mathrm{s}$ \cite{Palomba:1999su,Manchester:2004bp}. If the scalar field is coupled to the GJ electrons in the co-rotating magnetosphere or in the PC region, scalar dipole radiation from the pulsar could contribute to its spin-down. Scalar dipole radiation is only possible when the scalar mass $m_\phi\lesssim \Omega$. We write Eq. \ref{t3} as
\begin{equation}
\begin{split}
\frac{dE}{dt}=0.058~\mathrm{\frac{erg}{s}}\Big(\frac{g_e}{10^{-20}}\Big)^2 \Big(\frac{B_0}{8.5\times 10^{12}~\mathrm{G}}\Big)^2\Big(\frac{x}{10}\Big)^2 \Big(\frac{R}{14~\mathrm{km}}\Big)^8\Big(\frac{\Omega}{2\pi\times 29.56~\mathrm{Hz}}\Big)^6 \sin^2\theta_m\times \\
\Big(1-\frac{m^2_\phi}{\Omega^2}\Big)^\frac{3}{2},  
\label{use7}
\end{split}
\end{equation}
for co-rotating magnetosphere and 
\begin{equation}
    \begin{split}
        \frac{dE}{dt}=1.18\times 10^{-14}~\mathrm{\frac{erg}{s}}\Big(\frac{g_e}{10^{-20}}\Big)^2 \Big(\frac{B_0}{8.5\times 10^{12}~\mathrm{G}}\Big)^2\Big(\frac{h}{7~\mathrm{m}}\Big)^2 \Big(\frac{R}{14~\mathrm{km}}\Big)^8\Big(\frac{\Omega}{2\pi\times 29.56~\mathrm{Hz}}\Big)^8\\
        \sin^2\theta_m\times\Big(1-\frac{m^2_\phi}{\Omega^2}\Big)^\frac{3}{2},  
\label{use8}
    \end{split}
\end{equation}
for polar cap model. The contribution of scalar field radiation to the spin-down of the Crab pulsar is minimal in both models, as evident from Eqs. \ref{use7} and \ref{use8}. This contribution can be slightly enhanced for pulsars with stronger magnetic fields and higher spin frequencies. The scalar field contribution in spin-down is obtained for GJ electrons near the surface of the pulsar. The uncertainties in measuring the input parameters of the pulsar arise due to different EoS, timing models, intergalactic medium and parallax method. Given that the contribution of scalar radiation to pulsar spin-down is within the measurement uncertainty, we derive a bound on the scalar coupling as $g_e\lesssim 0.3$, using Eq. \ref{use7}. This bound is significantly weaker compared to all other existing constraints.  Therefore, we do not show the result in FIG. \ref{plotnew}. The net electrons within the star do not contribute to the pulsar spin-down since this charge is uniformly distributed throughout the star. To derive bounds on the scalar coupling related to the net charge in spin-down, there must be a charge imbalance in the star, where positive and negative charges accumulate at the two poles, forming an electric dipole.

\subsection{Effects of cosmic neutrino background on scalar couplings}
\label{subsece}
\begin{figure}
\centering
  \includegraphics[width=12cm]{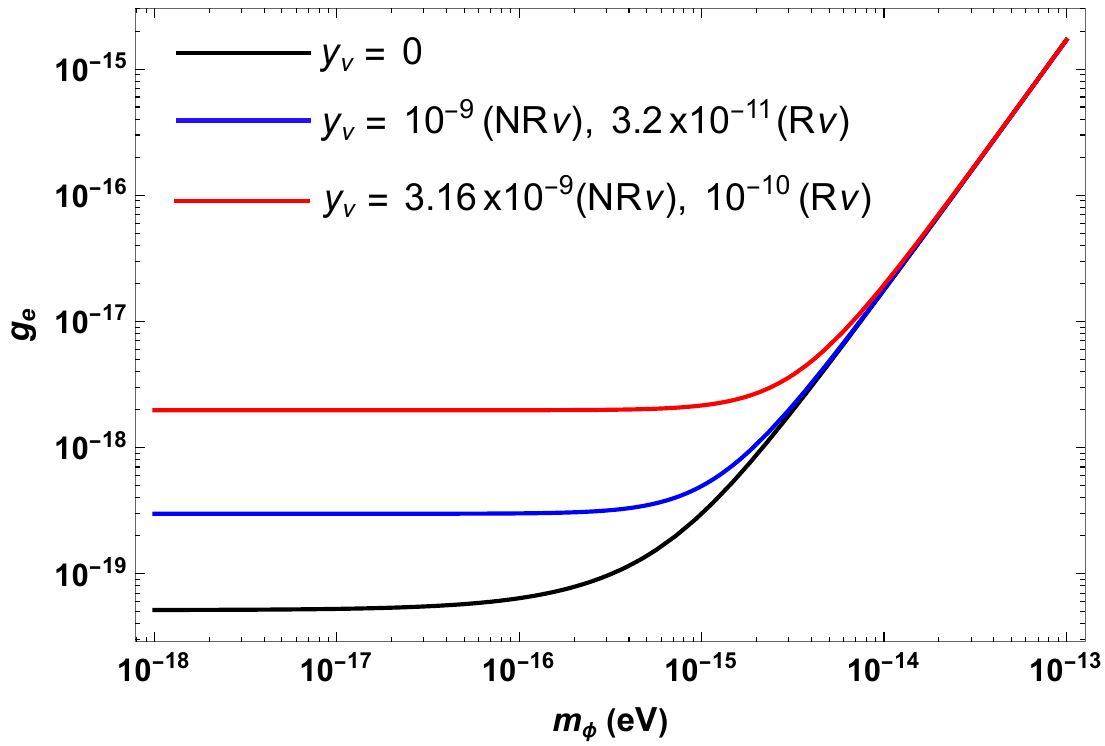}
  \caption{Limits on $g_e$ for various values of $y_\nu$ from the search for a long-range force.}
  \label{newplot2}
\end{figure}
In FIG. \ref{newplot2}, we present the constraints on the scalar-electron coupling when the scalar also interacts with neutrinos in the C$\nu$B. Using the PSR J0737-3039A/B double binary pulsar, we analyze the scalar-mediated long-range force between the two pulsars, where the scalar is sourced by the electrons within the star. Assuming the scalar interacts with cosmic neutrinos in the background, the scalar acquires a thermal mass from the C$\nu$B density, leading to a significant increase in the scalar mass for certain scalar-neutrino couplings. As the scalar mass increases, the force becomes effectively short-ranged. The C$\nu$B can be either relativistic (R) or non-relativistic (NR), depending on the neutrino mass. The slanted lines in the figure indicate the couplings derived in the non-zero scalar mass limit. For $y_\nu=0$ (black line), there is no C$\nu$B interaction, but as $y_\nu$ increases, the long-range force is suppressed, weakening the bound on $g_e$. The blue and red lines correspond to the increase in the scalar mass squared by $\Delta m^2_\phi\sim 10^{-30}~\mathrm{eV}^2$ and $\Delta m^2_\phi\sim 10^{-29}~\mathrm{eV}^2$ respectively. The enhanced mass squared values correspond to certain scalar-neutrino couplings as designated in the figure for both relativistic and non relativistic neutrinos. The shift of the bend point to higher masses indicates the increase in scalar mass. In FIG. \ref{newplot2}, we show the suppression of the bound on scalar-electron coupling derived from long-range force searches. Similarly, the constraint obtained from orbital period decay could also be weakened due to the influence of the C$\nu$B. The bound on the scalar coupling weakens as the value of $y_\nu$ increases. However, $y_\nu$ cannot increase indefinitely due to constraints from Big Bang Nucleosynthesis (BBN) and the effective number of neutrino species $N_{\mathrm{eff}}$. These constraints limit $y_\nu$ to $y_\nu\lesssim 10^{-5}$, as discussed in \cite{Huang:2017egl,Li:2023puz}. For instance, with $y_\nu\sim 10^{-6}$ for relativistic neutrinos, the correction to the scalar mass is approximately $\Delta m^2_\phi\sim 10^{-21}~\mathrm{eV}^2$, and the scalar-electron coupling is quenched to $g_e\lesssim 1.7\times 10^{-10}$.

\section{Conclusions and discussions}\label{sec8}
In this paper, we derive constraints on the scalar-electron coupling for ultralight scalar masses based on observations of the Crab pulsar and the double pulsar binary PSR J0737-3039A/B. We present bounds on the coupling from various methods, including the search for orbital period loss in double pulsar binaries, long-range force searches, and pulsar spin-down analysis. Our results are compared with existing limits on the scalar-electron coupling. The electrophilic scalar considered in our study is not required to be a DM candidate. Additionally, we explore the impact of the C$\nu$B on the obtained limits.

An ultralight scalar field can couple to electrons either within the pulsar magnetosphere or inside the pulsar itself, leading to a long-range field outside the pulsar magnetosphere or pulsar respectively. We calculate the scalar field profile both inside and outside the pulsar and its magnetosphere, depending on whether the field is sourced by net electrons within the pulsar or by Goldreich-Julian (GJ) electrons in the magnetosphere. The scalar field's mass is constrained by the pulsar's radius if sourced by net electrons, or by the light cylinder radius if sourced by GJ electrons. We find that the scalar field profile follows a Yukawa-like behavior, assuming constant density distributions for relativistic electrons in the magnetosphere and non-relativistic electrons within the star, which serve as reasonable approximations. Given the complexity of pulsar and magnetosphere dynamics and the absence of a definitive model to describe them, this simplified approach provides a framework for exploring electrophilic scalar couplings.

The long-range scalar field can also radiate from a binary pulsar system, like PSR J0737-3039A/B, and contribute to orbital period loss alongside GW radiation. Using pulsar timing data for the orbital period loss of this double pulsar system, we derive bounds on the scalar-electron coupling: $g_e \lesssim 7.24 \times 10^{-19}$ for GJ electrons and $g_e \lesssim 3 \times 10^{-21}$ for electrons within the star. These bounds are valid for scalar masses $m_\phi \lesssim 4.79 \times 10^{-19}~\mathrm{eV}$, constrained by the system's orbital frequency. The bound for the net electron charge inside the star is five orders of magnitude tighter than the WD limit and stricter than the fifth force bounds for $m_\phi \lesssim 10^{-21}~\mathrm{eV}$.

In the massless scalar limit, the Coulombic behavior of the scalar field profile generates an effective scalar-induced charge on the star or within the magnetosphere. This scalar charge leads to a long-range force between two stars in a binary system. Using PSR J0737-3039A/B as a test case to evaluate the scalar-mediated fifth force between the pulsars, we obtain constraints on the scalar-electron coupling: $g_e \lesssim 6.79 \times 10^{-14}$ for GJ electrons and $g_e \lesssim 5.14 \times 10^{-20}$ for electrons inside the star. These bounds are valid for scalar masses $m_\phi \lesssim 4.5 \times 10^{-16}~\mathrm{eV}$, constrained by the distance between the two pulsars in the binary.

We examine both the co-rotating magnetosphere model and the polar cap model to estimate the energy loss due to scalar emission and its contribution to the spin-down of the pulsar. Using the Crab pulsar as an example, we derive the bound on the scalar coupling based on this spin-down effect. Scalar radiation is only possible if the scalar mass is less than the pulsar's spin frequency given as $m_\phi\lesssim 1.22\times 10^{-13}~\mathrm{eV}$. However, in both models, we find that the contribution of scalar radiation to the pulsar's spin-down is minimal, leading to unpromising bounds on the scalar coupling. Our analysis assumes the density of GJ electrons, as the net electrons in the pulsar, being uniformly distributed, do not contribute to the spin-down. To induce a spin-down effect from the net charge, there would need to be a charge imbalance between the two poles of the star, creating an electric dipole.

The most stringent bound presented in this paper is derived from the orbital period decay of the double pulsar binary. This constraint is five orders of magnitude stronger than the limit set by white dwarf observations and also surpasses the fifth-force limit in the low scalar mass regime, $m_\phi\lesssim \mathcal{O}(10^{-21})~\mathrm{eV}$. However, the bound on scalar-electron coupling is four orders of magnitude weaker than the strongest existing limit established by the MICROSCOPE experiment.

Our bound on the coupling sourced by the GJ charge could become tighter with observations of compact stars exhibiting stronger magnetic fields, higher angular velocities, and with experiments of improved sensitivity. Similarly, the bounds arising from net charge can be enhanced with more precise measurements.

Currently, pulsar mass measurements achieve an accuracy of about $10^{-3}-10^{-4}$. If future observations improve this accuracy to $10^{-7}$ (for orbital period decay) and $10^{-15}$ (for long-range force detection), the derived bounds on scalar-electron coupling could potentially surpass the current MICROSCOPE limit.

Finally, we assess the impact of the C$\nu$B in deriving the scalar-electron coupling constraints from the search for long-range forces in the PSR J0737-3039A/B system. The C$\nu$B is omnipresent, and the long-range scalar field receives thermal corrections to its mass due to interactions with the C$\nu$B. For certain scalar-neutrino couplings, the scalar mass increases. As the scalar mass grows, the long-range force transitions into a short-range force, thereby weakening the bounds on the scalar-electron coupling. We find that for a non-zero scalar-neutrino coupling of approximately $y_\nu \sim 10^{-9}$ with non-relativistic C$\nu$B, the bound becomes $g_e \lesssim 3.22 \times 10^{-19}$. For relativistic C$\nu$B, with $y_\nu \sim 3.2 \times 10^{-11}$, the coupling is similarly constrained. The bound on the scalar-electron coupling due to the presence of C$\nu$B can also be weakened in other contexts, such as through searches for orbital period loss measurements.

Ultralight scalar particles can also be the promising candidates for DM, offering a solution to the core-cusp problem while avoiding constraints from direct detection experiments. Future experiments with enhanced sensitivity and compact stars with larger magnetic field and angular velocity could further tighten the bounds derived in this study. These results can also be extended to the radiation of ultralight vector or tensor particles from isolated pulsars or pulsar binary systems. Additionally, ultralight particles may couple with other leptonic currents, such as muons, allowing for the derivation of similar leptophilic bounds in those cases.

\section*{Acknowledgements}
The authors would like to thank Garv Chauhan and Anirudh Prabhu for useful discussions. G.L and T.K.P thank COST Actions COSMIC WISPers CA21106 and BridgeQG CA23130 supported by COST (European Cooperation in Science and Technology). This work is also supported by the ``QGSKY" project.

\bibliographystyle{utphys}
\bibliography{bira}
\end{document}